\documentclass[twocolumn,showpacs,aps,amsmath,amssymb,prb,preprintnumbers]{revtex4}
\newcommand{\BE}{\begin{equation}}
\newcommand{\EE}{\end{equation}}
\newcommand{\BA}{\begin{eqnarray}}
\newcommand{\EA}{\end{eqnarray}}
\usepackage{graphicx}
\usepackage{dcolumn}
\usepackage{bm}
\input epsf.tex

\begin{document}

\title{Directed diffraction without negative refraction}

\author{Hung-Ta Chien$^1$} \author{Hui-Ting Tang$^1$}\author{Chao-Hsien Kuo$^2$}\author{Chii-Chang Chen$^1$}\author{Zhen
Ye$^2$}\email{zhen@phy.ncu.edu.tw} \affiliation{$^1$Institute of
Optical Science, and $^2$Wave Phenomena Laboratory, Department of
Physics, National Central University, Chungli, Taiwan}

\date{December 22, 2003}

\begin{abstract}

Using the FDTD method, we investigate the electromagnetic
propagation in two-dimensional photonic crystals, formed by
parallel air cylinders in a dielectric medium. The corresponding
frequency band structure is computed using the standard plane-wave
expansion method. It is shown that within partial bandgaps, waves
tend to bend away from the forbidden directions. This phenomenon
perhaps need not be explained in terms of negative refraction or
`superlensing' behavior, contrast to what has been conjectured.

\end{abstract}

 \maketitle

Ever since the suggestion that a perfect lens can be realized by
so called Left Handed Materials (LHMs) or Negative Refraction
Index Materials (NRIMs)\cite{Pendry}, a conceptual material first
introduced by Veselago many years ago \cite{Ves}, the research on
superlenses and LHMs has been skyrocketing in the mist of much
debate. A great body of literature has been and continues to be
generated. The research ranges from finding characteristics of
LHMs, and negative refraction behavior (NRB), to fabricating
composite materials which may reveal as LHMs. All this can be
referred to in Refs.~\cite{webs}. In spite of some serious
debates\cite{Debates}, the common consensus seems to agree that
the negative refraction, perhaps equivalently the LHMs, has been
confirmed\cite{Pendry2}.

Up to date, there mainly three aspects regarding negative
refraction; this has caused some confusion to a certain degree.
First, the concept of negative refraction was invented and
specified with respect to LHMs, as originally proposed by
Veselago\cite{Ves}. For composite materials, the individual
components may be positively refractive. But when put together,
the materials as a whole may behave {\it effectively} as a
negative refracting medium, and a negative refraction index can be
defined\cite{Smith}. The second is to regard the refraction into
the same side as the incident wave as negative
refraction\cite{Zhang}. The second aspect can be observed for
waves across the boundary between an isotropic and an anisotropic
medium, and this is in fact irrelevant to the negative refraction
in its original sense\cite{Yau}. The third aspect is to show
negative refraction without employing or having to resort to an
effective index of negative refraction.

In a recent communication, Luo et al. \cite{PRB} described an
all-angle negative refraction by photonic crystals, without
negative effective index. By simulation, the authors demonstrated
a frequency range so that for all incident angles one obtains a
single negative-refracted beam. It was shown that a slab of
photonic crystal can focus a point source on one side of the slab
into a {\it real} point image on the other side even for the case
a parallel sided slab of materials, that is, the flat slab
imaging. This phenomenon has been connected to the superlensing
phenomenon discussed by Veselago\cite{Ves} and Pendry\cite{Pendry}
for a slab of LHMs. The simulation in Ref.~\cite{PRB} has
stimulated a number of further experimental and theoretical
attempts in verifying the negative refraction behavior of photonic
crystals \cite{Nature}.

Two problems, however, may concern with the superlensing
phenomenon and the negative refraction discussed in
Ref.~\cite{PRB}. The first question is whether the superlensing or
the flat slab imaging can be explained by other mechanisms. The
second question is whether there is indeed negative refraction as
suggested. The first question has been answered in Ref.~\cite{Ye}.
As shown in Ref.~\cite{Ye}, the imaging by a photonic crystal slab
can be caused by partial bandgap effects or anisotropic
scattering. The answer to the second question, in our view, is not
conclusive. In this communication, we wish to shed some light to
the second question, in the hope that more discussions can be
stimulated.

We will consider exactly the same two dimensional photonic crystal
systems as in Ref.~\cite{PRB}. Specifically, we consider a square
lattice of identical air cylinders (holes) in a dielectric medium
with dielectric constant $\epsilon = 12$. The lattice constant is
denoted by $a$ and the hole radius $r=0.35a$. In accordance with
Ref.~\cite{PRB}, we consider TE modes, that is, the magnetic field
is kept parallel to the axis of the air cylinders.

In order to compute the electromagnetic propagation through the
arrays of air cylinders, we have performed finite-difference
time-domain (FDTD) simulations with perfectly matched layer (PML)
boundary conditions. In the simulation, we scale all lengths by
the lattice constant and the angular frequency by $2\pi c/a$ to
make the system dimensionless.

First, we would like to duplicate the flat slab imaging observed
in Ref.~\cite{PRB}. The band structure of the system is plotted in
Fig.~1(a), in full agreement with the result in Fig.~3 of
Ref.~\cite{PRB}.

By the FDTD method, we have simulated the wave propagation across
a flat (11)-oriented slab of lattice of air cylinders. Hereafter,
the slabs of photonic crystals are all placed in air, in line with
the Ref.~\cite{PRB}. The geometric parameters are given in the
figure caption. Fig.~1(b) shows a snapshot of the magnetic field
for a point source transmitting the continuous-wave of frequency
0.192, which lies within the regime of so called all-angle
negative refraction (AANR)\cite{PRB}. Note that we have also done
with frequency 0.195 as in Ref.~\cite{PRB}, the results are
qualitatively the same. Here we indeed observe the formation of a
`point' image on the right hand side of the slab, confirming the
result depicted by Fig.~4 in Ref.~\cite{PRB}.

If such an imaging property can be attributed to the occurrence of
the superlensing effect (referring to the bottom paragraph on the
second page of Ref.~\cite{PRB}, two wave propagation schemes can
be identified. These are illustrated by Fig.~1(c) and (d)
respectively. Here, Fig.~1(c) shows the original sketch for the
superlensing. The superlensing effect will not only focus a single
image on the other side of the slab, but also another focused
image within the slab. These two images must go hand in hand. The
image inside the slab, however, is not obvious from Fig.~1(b). We
note that the imaging behavior has been misinterpreted in
Ref.~\cite{Bikash}. The second possible scheme is described by
Fig.~1(d). That is, a plane wave incident upon a slab of negative
refraction structures will be negatively refracted inside the slab
and recovers to its original travelling direction when returning
to the outside medium which is positive refracting.

While the point source scheme has been discussed in
Ref.~\cite{Ye}, here we will concentrate on the second scheme,
i.~e. Fig.~1(d). We wish to verify whether the negative refraction
depicted in Fig.~1(d) or asserted in the right panel of Fig.~2 in
Ref.~\cite{PRB} actually occurs for the waves within the AANR
regime, as suggested in Ref.~\cite{PRB}. For the purpose, we
consider electromagnetic propagation through much larger slabs and
plot the waves inside the slabs with high resolution, again, by
the standard FDTD method.

In Fig.~2, we plot the images of the magnetic intensity fields
when a collimated plane wave, mimicked by the Gaussian beam, is
incident at various angles onto a slab of the above photonic
crystals. Although many incident angles have been taken into
account, here we only show the results for four incident angles.
The slab measures as 49.5x92 in terms of lattice constant. The
frequency is 0.192, which is within the AANR range declared in
Ref.~\cite{PRB}. The flat slab is (11) oriented, as in the
simulation of Ref.~\cite{PRB}. The detailed information about the
setup can be referred in the figure and the figure caption. Here
we see that no matter what angle the wave is incident onto the
slab, the wave always tends to travel along the [11] direction of
the square lattice. This phenomenon has also been confirmed by
plotting the snapshots of the magnetic field. It was shown in
Ref.~\cite{Ye} that the [11] direction, in which there is a
passing band, serves as a main guiding channel for wave
propagation, and an focused image can be formed across the slab.
In the present case, the outgoing waves are more less parallel to
the incident waves. Here, however, we fail to observe the negative
refraction expected for the superlens as discussed in
Ref.~\cite{PRB} or that depicted in Fig.~1(d). We also verify with
other frequencies within the partial bandgaps, yielding similar
results.

We have also investigated electromagnetic propagation through a
slab of the photonic crystals with different orientations. As
example, we show the case with the [11] direction being tilted
leftward at 22.5 degree. The results are shown in Fig.~3. Here the
flat slab has the same dimension as that in Fig.~2. The normal
direction at the incident interface lies exactly in the middle of
the [11] and [10] directions of the square lattice of the air
cylinders. All the physical parameters are the same as in the
above and in Ref.~\cite{PRB}. Here we, again, observe that at any
incident angle, the waves are more or less bent towards the [11]
direction. In the case of Fig.~3(d), we also observed a secondary
path along the direction perpendicular to the indicated [11]
direction. This can be expected to be due to the symmetry of the
square lattice.

To exclude the boundary of the slab as a possible cause for the
bending, we have also performed simulations on various slab sizes.
Two examples are shown in Fig.~4: in (a) the slab height is
increased, while in (b) the slab horizontal width is enlarged. The
slab is (11) oriented. The plane wave is incident at the angle of
22.5 degree. Here, the waves inside the slab travel along the [11]
direction. Comparing the results in Figs.~2 and 4, one can
conclude that the wave deflection behavior inside the slab is not
caused by the boundaries.

All the results shown above indicate that in certain ranges of
frequencies, the photonic crystals can serve as a guiding medium
that directs the wave diffraction or deflection. The fact that the
waves can be bent into a certain direction may allow for novel
applications in controlling optical flows as suggested in
Ref.~\cite{Chen}.

One may ask for the reason for the discrepancies between the
present results and the analysis of the superlensing phenomenon
discussed in Ref.~\cite{PRB}. Here we would like to share our
thoughts. The details will be published elsewhere\cite{Kuo}. The
problem with the conjectured superlensing may lie in the approach
to the energy flow inside the slab. The usual approach mainly
relies on the curvatures of frequency bands to infer the energy
flow.

As documented in Ref.~\cite{Yeh}, an energy velocity is defined as
$\vec{v}_e = \frac{\frac{1}{V}\int \vec{J}_{\vec{K}}
d^3{r}}{\frac{1}{V}\int U_{\vec{K}} d^3{r}},$ where
$\vec{J}_{\vec{K}}$ and $U_{\vec{K}}$ are the energy flux and
energy density of the eigenmodes, and the integration is performed
in a unit cell. It can be shown that thus defined energy velocity
equals the group velocity obtained as $\vec{v}_g =
\nabla_{\vec{K}}\omega(\vec{K}).$ Therefore it is common to
calculate the group velocity to infer the energy velocity and
subsequently the energy flows or refraction of waves. It has been
discussed elsewhere\cite{Kuo}, whether the net current flow
through a unit cell really follows the direction of $\vec{v}_e$
remains unclear. We note here that the average flux through a
surface may be defined as $\langle\vec{J}\rangle =
\frac{\hat{n}}{S}\int d\vec{S}\cdot \vec{J}$, where $\hat{n}$ is
the unit normal vector of the surface $S$. Clearly, the volume
averaged current within a unit cell does not necessarily
correspond to actual energy flows.

In summary, we have used the standard FDTD method to simulate
electromagnetic propagation across various flat slabs of photonic
crystals. No negative refraction behavior has been discovered in
the frequency regime thought to be the regime for all angle
negative refraction. The present results, however, are consistent
with the previous simulation on the flat slab imaging by the
standard multiple scattering theory. Although the present analysis
has been focused on TM modes, it can be extended to TE modes. Some
results of TE modes have been reported in Ref.~\cite{Ye}.

The work received support from the National Science Council. Some
of the materials has been reported by one of us (ZY) at the
workshop on Left-Handed-Materials and Photonic Crystals in NCU
(December 10, 2003) and at the 5-th Pacific Rim Conference on
Lasers and Electro-Optics in Taipei (December 15-19, 2003).
Discussion with P.-G. Luan is thanked.

\section*{Figure Captions}

\begin{description}

\item[Fig. 1] (a) The band structure of a square lattice of air
holes embedded in the dielectric medium $\epsilon = 12$. The
lattice constant is $a$ and the radius of the cylinders is
$0.35a$. $\Gamma M$ and $\Gamma X$ denote the [11] and [10]
directions respectively. (b) The magnetic field $H_z$ of a point
source and its image across a photonic crystal slab called
superlens in Luo et al.\cite{PRB}. The slab measures 11.31x31.11,
and the frequency is 0.192. (c) The conceptual layout of a
superlensing phenomenon, adapted from Veselago\cite{Ves} and
Pendry\cite{Pendry}. (d) The illustration of negative refraction
of a plane wave across a flat slab, by analogy with the right
panel of Fig.~2 in Ref.~\cite{PRB}.

\item[Fig. 2] The intensity image of the magnetic fields across a
flat photonic crystal slab made of air holes in dielectric
$\epsilon = 12$. The principle directions of the crystal are drown
in the figure, as [11] and [10]. A plane wave is incident to the
slab with various incident angles in degrees with respect to the
[11] direction: (a) 0; (b) 22.5; (c) 45; and (d) 67.5. The slab
measures as 49.5x92 in terms of lattice constant. For clarity, the
air holes are not shown in the figure.

\item[Fig. 3] The intensity image of the magnetic fields across a
flat photonic crystal slab. The principle directions of the
crystal are drown in the figure, as [11] and [10]. The [11]
direction is tilted towards left by an angle of 22.5 degree with
reference to the normal direction of the slab surface. A plane
wave is incident to the slab with various incident angles in
degrees with respect to the normal direction of the slab surface:
(a) 0; (b) 22.5; (c) 45; and (d) 67.5. The slab measures as
49.5x92 in terms of lattice constant.

\item[Fig. 4] The intensity image of the magnetic fields across
two flat photonic crystal slabs. The principle directions of the
crystal are drown in the figure, as [11] and [10]. The two slabs
measure in terms of lattice constant as (a) 70x92;(b) 49.5x120. A
plane wave is incident to the slab with the incident angle of 22.5
degree with respect to the normal direction of the slab surface.

\end{description}

\end{document}